\documentclass[12pt]{article}
\usepackage{a4wide}
\usepackage{axodraw}
\usepackage{latexsym}
\usepackage{epsfig}
\usepackage{amsmath}
\usepackage{amssymb}

\def\as{\alpha_s}

\newcommand{\qq}{q\bar q}
\newcommand{\xt}{\hat x_T}

\newlength{\dinwidth} \newlength{\dinmargin}
\begin{document}

\thispagestyle{empty}

\begin{flushright}
  NIKHEF/2004-016\\
  ITP-UU-04/41\\
  Cavendish-HEP-2004/31\\
  DAMTP-2004-130\\
  hep-ph/0411241
\end{flushright}

\vspace{1.5cm}

\begin{center}
 {\Large\bf Joint resummation for heavy quark production}\\[1cm] 
  {\sc  Andrea Banfi$^{a,b,c}$, Eric Laenen$^{a,d}$\\
  \vspace{1.5cm}
  $^a${\it NIKHEF Theory Group\\
    Kruislaan 409, 1098 SJ Amsterdam, The Netherlands} \\[0.4cm]
  $^b${\it Cavendish Laboratory, University of Cambridge\\
    Madingley Road, CB3 0HE Cambridge, UK} \\[0.4cm]
  $^c${\it DAMTP, Centre for Mathematical Sciences\\
    Wilberforce Road, CB3 0WA Cambridge, UK} \\[0.4cm]
  $^d${\it Institute for Theoretical Physics, Utrecht University\\
    Leuvenlaan 4, 3584 CE Utrecht, The Netherlands} \\
}
\end{center}

\vspace{2cm}


\begin{abstract}

\noindent{
  We present joint threshold and recoil resummed transverse momentum
  distributions for heavy quark hadroproduction, at next-to-leading
  logarithmic accuracy. We study the dependence of these
  distributions on the production
  channel, the color configurations and the differences with the pure
  threshold-resummed distribution.  }

\end{abstract}

\vspace*{\fill}

\newpage
\reversemarginpar

\section{Introduction}
\label{sec:intro}

The formalism \cite{Li:1998is,Laenen:2000ij} of hadronic cross
sections for the joint resummation of distributions singular at
partonic threshold and at zero recoil has so far been applied to only
a few processes. The most recent studies involve processes that
proceed at lowest order through a $2 \rightarrow 1$ electroweak ($Z/W$
production \cite{Kulesza:2002rh}) or Yukawa interaction (Higgs
production \cite{Kulesza:2003wn}).  For these cases, the observables
are the production cross sections at fixed mass $Q$ and measured
$Q_T$. Partonic threshold is defined by $z\equiv Q^2/\hat{s} = 1$,
where $\hat{s}$ is the partonic center of mass energy squared, and
zero recoil by $Q_T=0$. At any finite order, the distributions take
the form of plus-distributions $\left[\ln^{k}(1-z)/(1-z)\right]_+$ and
$\left[\ln^{k}(Q/Q_T)/Q_T\right]_+$. In these observables the latter
distributions enter in the physical cross sections, whereas the former
are defined, after factorization, in the context of a perturbative
analysis of the hard scattering.

The basis for the joint threshold and recoil resummation for these
processes and those proceeding at lowest order through a $2\rightarrow
2$ (QCD) reaction was given in Ref.~\cite{Laenen:2000ij}. The first
application of the formalism to the specific case of prompt photon
hadroproduction was presented in Ref.~\cite{Laenen:2000de}.  There the
photon transverse momentum spectrum was analyzed, and a preliminary
numerical study performed. Recoil corrections were found to have a
significant impact on the photon spectrum, both at large and at small
values of $p_T$. 
For such $2\rightarrow 2$ processes,
the formalism implements the notion that, in
the presence of QCD radiation, the actual transverse momentum produced
by the hard collision is not $\vec{p}_T$ but rather $\vec p_T-\vec
Q_T/2$, with $\vec Q_T$ the total transverse momentum of soft
recoiling partons. In fact, the joint-resummed partonic $p_T$ spectrum
has the form of a hard scattering cross section as a function of
$p_T'\equiv|\vec p_T-\vec Q_T/2|$, convoluted with a {\em
  perturbative}, albeit resummed $\vec Q_T$ distribution.  The extreme
situation $Q_T = 2 p_T$ in which all transverse momentum is produced
through recoil leads to a singularity in the hard scattering function,
equivalent to the singularity at zero $p_T$ in the prompt-photon Born
cross section. We note however that a recently proposed 
extension \cite{Sterman:2004yk} of joint resummation avoids this 
singularity.

In this paper we apply the joint resummation formalism 
as given in Ref.~\cite{Laenen:2000ij} to another
prominent $2\to 2$ scattering observable: the $p_T$ distribution of
heavy quarks produced in hadronic collisions.  Key differences with
the prompt-photon case are, first, the presence of the heavy quark
mass $m$, preventing a singularity in the hard scattering function
when $Q_T = 2 p_T$ and, second, the possibility of multiple
colored states for the produced top quark pair.  We derive
joint-resummed and threshold-resummed transverse momentum distributions at
next-to-leading logarithmic (NLL) accuracy, and study for different
production channels the dependence on color configurations, as well as
differences among the two resummations and the exact next-to-leading
order (NLO) distribution. 

Other recent studies for heavy quark $p_T$ distributions based on
finite order expansions of threshold resummation 
can be found in Refs.~\cite{Kidonakis:2004qe,Kidonakis:2003vs}. 

Relevant formulas are derived in
section 2.  In section 3 we present numerical studies of the
dependence of the joint resummation effects on the production channel,
on color configurations and on the heavy quark flavor.  We also assess
the differences with the pure threshold-resummed distribution.
Section 4 contains our conclusions.

\section{The observable}
\label{sec:obs}

We consider the inclusive $p_T$ distribution of a heavy quark produced
via the strong interaction in a hadron-hadron collision at center of
mass (cm) energy $\sqrt{S}$
\begin{equation}
  \label{eq:el-proc}
  h_A(p_A) + h_B(p_B) \to Q(p_c)+X \>,
\end{equation}
where $h_{A,B}$ refers to the two incoming hadrons, $Q$ to the
detected heavy quark and $X$ to the unobserved part of the final state
which includes also the heavy anti-quark $\bar Q$.  The lowest order
QCD processes producing a heavy quark are
\begin{equation}
  \label{eq:parton-proc}
  \begin{split}
    &q(p_a) + \bar q(p_b) \to Q(p_c) + \bar Q(p_d)\>, \\
    &g(p_a) +  g(p_b) \to Q(p_c) + \bar Q(p_d)\>,
  \end{split}
\end{equation}
at cm energy $\sqrt{\hat{s}} = \sqrt{\xi_a\xi_b S}$ with $\xi_a,\xi_b$
parton momentum fractions.  Exact higher order corrections to the
differential cross sections for these partonic
processes have been computed to NLO
\cite{Nason:1989zy,Beenakker:1989bq,Beenakker:1991ma,Mangano:1992jk}.
To any order \cite{Collins:1989gx} the observable may be written in
the following factorized form (up to power corrections behaving as
$1/p^2_T$)
\begin{equation}
  \label{eq:crs-def}
  \frac{d\sigma_{AB\to Q+X}}{d p_T}=
  \sum_{a,b}\int_0^1 d\xi_a d\xi_b \, \phi_{a/A}(\xi_a,\mu)\phi_{b/B}(\xi_b,\mu)
  \frac{d\hat\sigma_{ab\to Q+X}}{dp_T}(\xi_a,\xi_b,\alpha_s(\mu),p_T)\>,
\end{equation}
with $d\hat\sigma_{ab\to Q+X}/dp_T$ the partonic differential
cross-section, $\phi_{a/A}$ and $\phi_{b/B}$ parton densities, and
$\mu$ the factorization and renormalization scale.  The purpose of this
study is to investigate certain effects of soft gluons in these higher order
corrections for the heavy quark $p_T$ distribution.

The first of these are threshold enhancements, which essentially
involve the energy of soft gluons.  In the context of the
factorization \eqref{eq:crs-def} we define hadronic and partonic
threshold by the conditions $S = 4m_T^2$ and $\hat{s} = 4m_T^2$,
respectively, with $m_T$ the transverse mass $\sqrt{m^2+p_T^2}$.  It
is convenient to define the scaling variables
\begin{equation}
  \label{eq:xT}
  x_T^2=\frac{4 m_T^2}{S}\>,\qquad
  \hat x_T^2=\frac{4 m_T^2}{\xi_a \xi_b S}\>\,,
\end{equation}
so that hadronic (partonic) threshold is at $x_T^2 =1$ ($\hat x_T^2 =
1$).  The higher order corrections to the partonic cross section
$d\hat\sigma_{ab}/dp_T$ contain distributions that are singular at
partonic threshold. Threshold resummation organizes such distributions
to all orders.

The second of these are recoil effects, resulting from radiation of
soft gluons from initial-state partons.  We wish to treat these
effects in the context of joint threshold and recoil resummation. To
this end we employ the refactorization analysis of
Refs.~\cite{Laenen:2000ij,Laenen:2000de}, which enables us to
identify a hard scattering with reduced cm
energy squared $Q^2$ and at transverse momentum $\vec{Q}_T$ with
respect to the hadronic cm system. This hard scattering produces a
heavy quark with transverse momentum
\begin{equation}
  \label{eq:transverse}
  \vec{p\,}_T'\equiv\vec p_T-\frac{\vec Q_T}{2}\>.
\end{equation}
The kinematically allowed range for the invariant mass $Q$ of the
heavy quark pair in this hard scattering is limited from below by
$2m_T' = 2\sqrt{m^2 + p_T'^2}$ so that threshold in the context of
joint resummation is defined by
\begin{equation}
  \label{eq:4}
    \tilde x_T^2\equiv\frac{4 m_T'^2}{Q^2} = 1\>.
\end{equation}
A refactorization analysis on the lines of the one performed in
Ref.~\cite{Laenen:2000ij} leads to the following expression for the
observable in Eq.~\eqref{eq:crs-def}
\begin{equation}
\label{eq:8}
    \frac{d\sigma_{AB\to Q+X}}{dp_T}
 = \int d^2 Q_T dQ^2\, \theta(\bar{\mu}-|\vec{Q}_T|)
\frac{d\sigma_{AB\to  Q+X}}{dp_T dQ^2 d^2 \vec Q_T}\,,
\end{equation}
where
\begin{equation}
  \label{eq:refactor}
  \begin{split} 
    \frac{d\sigma_{AB\to Q+X}}{dp_T dQ^2 d^2 \vec
      Q_T}&=\sum_{ab=q\bar{q},gg}\int d\xi_a d\xi_b
    \, \phi_{a/A}(\xi_a,\mu)\, \phi_{b/B}(\xi_b,\mu) \\
    & \hspace{-1cm} \times \int_0^{\xi_a} dx_a d^2 k_a\,
    \bar c_{a/a}\left(\frac{x_a}{\xi_a},\vec k_a\right) \int_0^{\xi_b}
    dx_b d^2 k_{b}\,
    \bar c_{b/b}\left(\frac{x_b}{\xi_b},\vec k_b\right)\\
    & \hspace{-1cm}\times \frac{dp_T'}{dp_T} \int_0^1 dw_s \>
    \left\{C_{\delta,\mathbf{1}}^{ab}(\alpha_s(\mu), \tilde x_T^2)
      \frac{d\sigma^{(0)}_{ab,\mathbf{1}}}{dp'_T}\, S'_\mathbf{1}(w_s)
      + C_{\delta,\mathbf{8}}^{ab}(\alpha_s(\mu), \tilde x_T^2)
      \frac{d\sigma^{(0)}_{ab,\mathbf{8}}}{dp'_T}\, S'_\mathbf{8}(w_s)\right\}\\
    & \hspace{-1cm} \times \frac{1}{S}\delta(1-Q^2/S-(1-x_a)-(1-
    x_b)-w_s) \delta(\vec Q_T+\vec k_a+\vec k_b)
  \end{split}
\end{equation}
is known as the profile function.  Note that we restrict ourselves to
relatively soft, perturbative recoil contributions by integrating over
$\vec Q_T$ in Eq.~\eqref{eq:8} such that $|\vec{Q}_T|\le \bar \mu$.
Implementing a cut-off $\bar\mu$ is of course a rather crude
approximation to what should be a consistent matching to finite order
results. The implementation of such a matching procedure is however
beyond the scope of this work.\footnote{Note that such a matching in
  the extension of Ref.~\cite{Sterman:2004yk} is straightforward.}  We
shall discuss the limitations on the integration variables such as
$\vec{Q}_T$ further in the next section.

Let us comment on the right hand side of Eq.~\eqref{eq:refactor},
moving from bottom to top. The last line implements transverse
momentum conservation, as well as the decomposition of the
(normalized) above-threshold energy $1-Q^2/S$ into contributions from
collinear radiation ($1-x_{a,b}$) and from wide-angle soft radiation
($w_s$). The definition of the observable and the refactorization
ensure \cite{Laenen:2000ij} that only the initial state contributes to
the recoil of the hard scattering.  The line above the last displays
the hard scattering functions $C_\delta \, d\sigma^{(0)}/dp_T'$, one
per color state (singlet or octet) in which the heavy quark pair can
be produced.  In a diagrammatic representation all its lines are
off-shell by at least an amount $p_T'$. The factor $dp_T'/dp_T$
accounts for the phase space difference between the hard scattering in
the factorized expression Eq.~\eqref{eq:crs-def}
($d\sigma^{(0)}/dp_T$) and here, where $d\sigma^{(0)}/dp_T'$ is
appropriate.  The functions $S'_\mathbf{1}$ and $S'_\mathbf{8}$
summarize the effects of coherent wide-angle soft radiation, and are
sensitive to the color structure of the hard scattering.  In
particular, the effect of soft radiation for our observable is such
that, to NLL accuracy, there is no mixing between singlet and octet
color states, as was already observed for the case of the total
heavy-quark cross section near threshold~\cite{Bonciani:1998vc}.

Moving up further in Eq.~\eqref{eq:refactor}, the functions
$\bar c_{a/a}$ are defined in Ref.~\cite{Laenen:2000ij}, and
are related to the density of parton $a$ in a parton of the same flavor at
fixed energy fraction $x_a$ and transverse momentum $\vec k_a$.  The
prime on the $S$ function indicates that factors
$\sqrt{U_{DY}}$, the square root of the Drell-Yan soft function, have
been introduced in equal numbers in numerator and denominator and then
conveniently redistributed between the heavy quark soft functions and
the $\bar c$ functions in order to remove gauge dependence in the
individual functions.  Finally the top line displays the convolution
with standard parton distribution functions, as well as the sum over
initial state parton flavors, in which the $g q,\, g\bar{q}$ channels
have been neglected because they are suppressed near threshold.

A convenient way to implement the energy and transverse momentum
conservation is by passing to transform space, where the impact vector
$\vec b$ is Fourier conjugate to $\vec Q_T$, and the variable $N$ is
Mellin conjugate to the threshold variables in Eqns.~\eqref{eq:xT} and
\eqref{eq:4}. This converts the convolution in Eq.~\eqref{eq:refactor}
to a product. We aim at NLL accuracy in transform space, that is we
account for all contributions $\exp\{\as^n L^{n+1}\}$ and $\exp\{\as^n
L^n\}$, where $L$ can be either $\ln N$ or $\ln b$.

Using their renormalization group scaling properties, moments of the
soft functions may, to next-to-leading logarithmic (NLL) accuracy, be
written as \cite{Kidonakis:1996aq,Kidonakis:1997gm}
\begin{equation}
  \label{eq:7}
  \tilde S'_\mathbf{1/8}\left(\frac{Q}{N\mu},\as(\mu)\right)  = 
  \tilde S'_\mathbf{1/8}\left(1,\as(Q/N)\right)
  \exp\left[ \int_{\mu}^{Q/N} \!\frac{d\mu'}{\mu'}\, 2\,\mathrm{Re}\,\Gamma_\mathbf{1/8}(\as(\mu'))  \right]\,.
\end{equation}
The solution needs careful definition in order to avoid values of $N$
for which $\alpha_s(Q/N)$ is singular. The $N$-integration contour we
choose indeed assures this, and anyhow, at NLL accuracy, we can
approximate the functions $ \tilde S'_\mathbf{1/8}(1,\alpha_s)$ by
their lowest order expressions.  The form of the one-loop soft
anomalous dimensions $\Gamma_\mathbf{1}$ and $\Gamma_\mathbf{8}$ we
need is straightforwardly derived from the expressions in
Ref.~\cite{Kidonakis:1996aq,Kidonakis:1997gm,Kidonakis:2001nj}. Their
real parts are
\begin{eqnarray}
  \label{eq:soft-anomalous}
    2\,\mathrm{Re}\,\Gamma_\mathbf{1}(\as) & = & -\frac{\as}{\pi}\,2C_F\left(\mathrm{Re} L_\beta+1\right)\,,\\ \label{eq:soft-anomalous-2}
    2\,\mathrm{Re}\,\Gamma_\mathbf{8}(\as) & = &
    2\,\mathrm{Re}\,\Gamma_\mathbf{1}(\as)
    +\frac{\as}{\pi}\,C_A\left(\ln\frac{m_T^2}{m^2}+ \mathrm{Re}
      L_\beta\right)\,,
\end{eqnarray}
\begin{equation}
  \label{eq:L-beta}
  \mathrm{Re} L_\beta = \frac{1+\beta^2}{2\beta}
  \left(\ln\frac{1-\beta}{1+\beta}\right)
  \>,
  \quad \beta=\sqrt{1-m^2/m_T^2} \>.
\end{equation}
We postpone a discussion of these expressions to further below.

To NLL accuracy we may approximate both $dp_T'/dp_T$ and the
$C_\delta$ functions by $1$, so that integrating over $Q^2$, and, with
manipulations similar to those of Ref.~\cite{Laenen:2000ij} we arrive
at
\begin{equation}
\label{eq:2}
  \begin{split}
    &\frac{d\sigma_{AB\to Q+X}}{dp_T d^2 \vec Q_T}= \sum_{ab=q\bar{q},gg}
    p_T \int\frac{d^2 b}{(2\pi)^2}e^{i \vec b \cdot  \vec Q_T}
    \int\frac{dN}{2\pi i} \phi_{a/A}(N,\mu)\>\phi_{b/B}(N,\mu)\>
    e^{E_{ab}(N,b)}\\
    &\frac{e^{-2\>C_F \>t(N)\>(\mathrm{Re} L_\beta+1)}}{4\pi S^2}\left(
      \tilde M^2_{\bf 1}(N)+\tilde M^2_{\bf 8}(N)
    e^{C_A \>t(N)\> \left(\ln \frac{m_T^2}{m^2}+L_\beta\right)}
    \right) 
        \left(\frac{S}{4(m^2+|\vec p_T-\vec Q_T/2|^2)}\right)^{N+1}
    \!\!\!\!\>.  
  \end{split}
\end{equation}
where the inverse Fourier transform and Mellin transform are explicit.
Notice in particular the last factor, which provides a kinematic link
between recoil and threshold effects.  The exponential functions
$E_{ab}$ \cite{Laenen:2000ij,Laenen:2000de} to next-to-leading
logarithmic (NLL) accuracy are
\begin{equation}
  \label{eq:Eab}
  E_{ab}(N,b) = \int_{\chi(N,b)}^Q \frac{d \mu'}{\mu'}
  [A_a(\as(\mu'))+A_b(\as(\mu'))] 2\ln\frac{\bar N \mu'}{Q} -g b^2
  \,,\qquad \bar N = N e^{\gamma_E}\,,
\end{equation}
where the coefficients $A_a$ and $A_b$ are taken from
Ref.~\cite{Laenen:2000ij}. We also added to the perturbative exponent 
the non-perturbative (NP) Gaussian smearing term $-gb^2$.

To simplify notations, we have combined in Eq.~(\ref{eq:2}) $\tilde
S'_\mathbf{1},\tilde S'_\mathbf{8}$ with the Born matrix elements that
contribute to $d\sigma^{(0)}/dp_T$ to build up the two Born functions
$\tilde M^2_{\bf 1}(N)$ and $ \tilde M^2_{\bf 8}(N)$, and we have
introduced the evolution variable
 \begin{equation}
   t(N)=\int_Q^{Q/N}\!\frac{d\mu'}{\mu'}\frac{\alpha_s(\mu')}{\pi}\>.
\end{equation}
The Born functions $\tilde M^2_{\bf 1}(N), \tilde M^2_{\bf 8}(N)$ are
the Mellin moments of the lowest order heavy quark production matrix
elements for either the $q\bar{q}$ or $gg$ channel, the index
labeling the color-state of the heavy quark pair:
  \begin{multline}\label{eq:14}
    \tilde M^2_{\qq, \bf 8}(N)\equiv\int_0^1 d\xt^2
    \frac{(\xt^2)^N}{\sqrt{1-\xt^2}}M^2_{\qq, \bf 8}(\hat x_T^2)\\
    = 16 \pi^2 \as^2\frac{C_F
      T_R}{N_c}\frac{\sqrt{\pi}\Gamma(N+1)}{\Gamma(N+5/2)}
    \left(N+2+(N+1)\frac{m^2}{m_T^2}\right)\>,
  \end{multline}
  \begin{multline}\label{eq:5}
    \tilde M^2_{gg, \bf 1}(N)\equiv \int_0^1 d\xt^2
    \frac{(\xt^2)^N}{\sqrt{1-\xt^2}}M^2_{gg, \bf 1}(\hat x_T^2)\\
    = 16 \pi^2 \as^2\frac{2T_R}{N_c(N_c^2-1)}\frac{\sqrt{\pi}
      \Gamma(N+1)}{\Gamma(N+3/2)}
    \left(\frac{N+1}{N}+2\left(\frac{m^2}{m_T^2}-\frac{m^4}{m_T^4}\right)
    \right)\>,
    \end{multline}
  \begin{multline}\label{eq:6}
    \tilde M^2_{gg, \bf 8}(N)\equiv \int_0^1 d\xt^2
    \frac{(\xt^2)^N}{\sqrt{1-\xt^2}}M^2_{gg, \bf 8}(\hat x_T^2)\\ =
    \frac{ 16 \pi^2
      \as^2}{N_c^2-1}\frac{\sqrt{\pi}\Gamma(N+1)}{\Gamma(N+3/2)}
    \left[ \left(2C_F-\frac{1}{N_c}\right)
      \left(\frac{N+1}{N}+2\left(\frac{m^2}{m_T^2}-\frac{m^4}{m_T^4}\right)
      \right)\right.\\ \left.  -\frac{C_A}{2N+3} \left(\frac{N+1}{N}+
        2(N+1)\left(\frac{m^2}{m_T^2}-\frac{m^4}{m_T^4}\right)\right)\right]\>,
  \end{multline}
  with $M^2_{\bf 1}$ and $M^2_{\bf 8}$ the square matrix element for
  the production of a hard $Q\bar Q$ system in a singlet and octet
  color state respectively. We point out that, at this order, in the
  $q\bar q$ channel the heavy quark-antiquark pair is produced only in
  an octet state.

In a refactorization analysis, the expressions
  $2\mathrm{Re}\,\Gamma_\mathbf{1}$ and
  $2\mathrm{Re}\,\Gamma_\mathbf{8}$ are related to the (real parts of)
  anomalous dimensions of certain operators, composed of Wilson lines
  \cite{Kidonakis:1997gm,Kidonakis:1998nf,Kidonakis:1998bk}, as the
  structure of Eqs.~\eqref{eq:7} reflects.  Because for our case
  imaginary parts of the virtual corrections (Coulomb phases) to the
  anomalous dimension cancel in the cross section, the soft function
  can be given a natural probabilistic interpretation. 

  From this viewpoint, at NLL accuracy, soft wide-angle gluons can be
  considered as being emitted independently from the external partons,
  and approaching threshold is equivalent to suppressing all soft
  gluon emissions with energies above $Q/N$.  Furthermore, due to the
  fact that there is no mixing between different color structures,
  soft gluon emission exponentiates straightforwardly.  The
  exponential in Eq.~\eqref{eq:7} can then be interpreted as the
  probability of not emitting any wide-angle soft gluon with energy
  above $Q/N$.
  We now offer some comments on the explicit expression of the anomalous
  dimensions in some limiting cases.
\begin{itemize}
\item In the case in which the heavy quarks are at rest in their cm
  frame, i.e. $\beta \to 0$, $\mathrm{Re} L_\beta \to -1$, and
  $2\mathrm{Re}\Gamma_\mathbf{1}$ vanishes, corresponding to the fact
  that the combined heavy quark-anti-quark pair forms a color-singlet.
  On the other hand, $2\mathrm{Re}\,\Gamma_\mathbf{8} \to -\as
  C_A/\pi$, which reflects the fact that the quark-antiquark pair
  produced in an octet state radiates as a single object rather than
  two separate particles.
\item For positive moderate values of $\beta$ ($0 < \beta < 1$) we
  observe that $\Gamma_\mathbf{8}$ contains the term
\begin{equation}
  \label{eq:13}
 \ln\left(\frac{m_T^2}{m^2} \right)
= \ln\left(\frac{(2 p_a\cdot p_c)\, (2 p_b \cdot p_c)}{(2p_a\cdot p_b) \, m^2} \right)\,,
\end{equation}
which is the standard contribution to the anomalous dimension from the
$q\bar q g$ antenna.
\item In the extreme region $p_T \gg m$ ($\beta
  \rightarrow 1$),  gluons can become effectively collinear
  to the heavy quarks. In this case $L_\beta$ approximates $\ln
  p_T^2/m^2$, a large {\em collinear} logarithm. The resulting
  logarithmic contributions are identical in $\Gamma_\mathbf{1}$ and
  $\Gamma_\mathbf{8}$, reflecting the fact that collinear radiation
  depends only on the color charge of each hard emitting parton, and is not
  sensitive to the color or geometrical structure of the event. In
  this large $p_T$ limit, collinear logarithms should be resummed to
  all orders, and a different analysis is required
  \cite{Cacciari:1998it,Cacciari:1994mq}.
\end{itemize}

This completes our discussion of the joint-resummed transverse
momentum distribution.
The threshold-resummed result can now easily be derived, by
substituting Eq.~\eqref{eq:2} into \eqref{eq:8} (after the performing
the $Q^2$ integral) and neglecting $\vec Q_T$ in the last factor in
Eq.~\eqref{eq:2}. Then the $\vec Q_T$ integral sets $\vec b$ to zero
everywhere, yielding the threshold-resummed result.

\section{Top and bottom quark $p_T$ spectra}
\label{sec:top-bottom-quark-1}

In this section we demonstrate numerically the soft gluon effects for
top and bottom quark transverse momentum distributions for collider
and fixed target kinematics respectively. Rather than performing a
detailed comparison with data, our aim is to examine the numerical
behavior of our formulas in realistic settings. We leave therefore also
to a future study the application of our results for the description of
charm quark/charmed meson $p_T$ spectra in fixed target kinematics.

Let us first comment on the allowed integration range of the recoil
momentum $Q_T$. In Eq.~\eqref{eq:8} we have simply imposed an upper
limit $\bar\mu$ on the modulus of this transverse momentum vector.
There is however another constraint on $Q_T$, because the recoil
factor at the end of Eq.~\eqref{eq:2} implies the condition
\begin{equation}
  \label{eq:1}
   \left(\frac{S}{4(m^2+|\vec p_T-\vec Q_T/2|^2)}\right) > 1\,,
\end{equation}
 which expresses the fact that the available hadronic energy
 be more than the minimum mass of the final state in the recoiling 
 hard scattering.
Clearly, for the maximum value $p_T^2= S/4-m^2$,
$\vec{Q}_T$ has no phase space left. For a given value $p_T$ and
(large) $\bar\mu$ the allowed range can be chosen to be the overlap of
the two disks $D_{1,2}$ in $\vec Q_T$ space
\begin{equation}
  \label{eq:3}
  D_1:\quad |\vec Q_T - 2 \vec p_T |^2 \leq S - 4m^2,\quad
  D_2:\quad |\vec Q_T | \leq \bar\mu \>.
\end{equation}
or one can choose $\bar\mu$ such that disk $D_2$ fits in $D_1$.  We
chose the former option. Notice that due to the heavy quark mass, the
expression in Eq.~\eqref{eq:2} stays finite at $Q_T = 2p_T$ , in
contrast to the prompt photon production case analyzed in
Refs.~\cite{Laenen:2000ij,Laenen:2000de}.

For top quark production we chose the cut-off $\bar{\mu} = 200$ GeV.
This might seem quite large as a cut-off on recoil momentum, given that the
joint-resummation formalism assumes that recoil radiation is soft.
This large value is however relatively innocuous, because, as we will
see, for the joint-resummed cross section the dominant $Q_T$ values
are quite moderate, typically a few percent of the hard scale. The
benefit of this large cut-off value is that, while recoil effects are
still dominated by soft gluons, it allows us to recover numerically
the threshold-resummed result, in the manner described at the end of
section \ref{sec:obs}.  For bottom quark production we choose
$\bar{\mu} = 30$~GeV, the motivation for this large value being
similar to that for the top quark.

Before we show results, let us state our default choices for various
input parameters. For these and other plots we use the NLO GRV parton
density set of Ref.~\cite{Gluck:1998xa} corresponding to
$\alpha_s(M_Z)=0.114$, with the evolution code of
Ref.~\cite{Vogt:2004ns}, changing flavor number at $\mu = m_c (1.4
\,\mathrm{GeV}),\; m_b (4.5 \, \mathrm{GeV})$.  We chose the
factorization and renormalization scale equal to the transverse mass $m_T$.
For the non-perturbative Gaussian smearing parameter $g$ in
Eq.~\eqref{eq:Eab} we took $g=1\,\mathrm{GeV}^2$.  Furthermore, to
avoid spurious singular behavior at the subleading level, we used,
following \cite{Kulesza:2002rh}
\begin{equation}
  \label{eq:9}
\chi(bQ,N) =  \bar{b} + \frac{\bar N}{1+\eta \bar{b}/\bar{N}}\,,
\qquad \bar{b} = \frac{b Q\, e^{\gamma_E}}{2}\,,
\end{equation}
with $\eta = 1/4$. 

\begin{figure}[htbp]
  \centering
  \epsfig{file=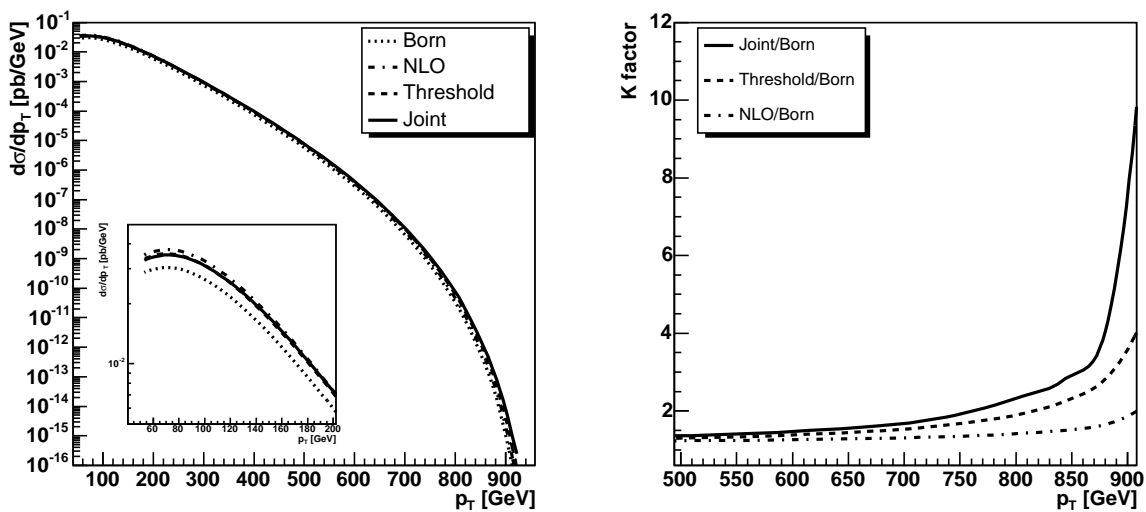,width=\textwidth}
  \caption{Top quark $p_T$ spectra and 
    $K$-factors for the $q\bar q$ channel.  \label{fig:ptspectra_qq}}
\end{figure}

We begin with top quark ($m=178$ GeV) production, and compare in Figure \ref{fig:ptspectra_qq}
joint-resummed, threshold-resummed and the exact LO and NLO calculations
for the top quark $p_T$ spectrum produced in the $q\bar q$ channel 
at the Tevatron ($\sqrt{S} = 1.96 \, \mathrm{TeV}$).
We observe that, while the resummed and NLO curves are close 
for small and moderate
$p_T$ (the inset provides a somewhat better view of the low $p_T$ region), 
for large $p_T$ values the resummed curves depart significantly
from the NLO curve.  Of course, cross sections for top quark
production at such large $p_T$ at the Tevatron are far too small to be
measured, so that our plots at large $p_T$ have only theoretical
interest.  For such large $p_T$ values, the hadronic threshold,
defined in Eq.~\eqref{eq:xT}, approaches the partonic one, where
larger $N$ values dominate, a prerequisite for seeing significant
effects for both resummations.  The enhancements relative to the Born
cross section are shown in the form of a K-factor\footnote{Normalized
  to the Born cross section.}  in Figure~\ref{fig:ptspectra_qq}.  As
expected, threshold resummation produces an overall enhancement of the
cross section that increases with increasing $p_T$, yielding e.g. a
$35\%$ enhancement over NLO at $p_T = 800\,\mathrm{GeV}$. Joint resummation almost
doubles that effect, which can be understood from the arguments given
in the introduction:  the joint-resummed enhancement at large $p_T$
effectively constitutes a smearing of the threshold-resummed $p_T$ spectrum by
a resummed recoil function.
\begin{figure}[htbp]
  \centering
  \epsfig{file=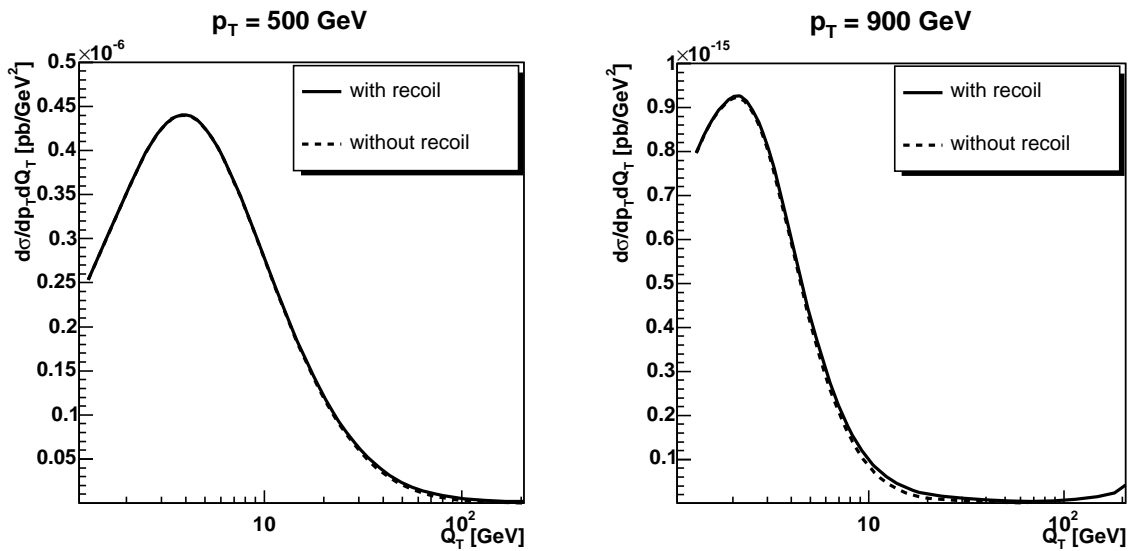,width=\textwidth}
  \caption{Top quark profile functions (Eq.~\eqref{eq:2})
 for the $q\bar q$ channel, at $p_T = 500$ and  $900$ GeV. \label{fig:profile_qq}}
\end{figure}

Before showing results for the $gg$ channel, we exhibit in
Fig.~\ref{fig:profile_qq} the $Q_T$ profile of Eq.~\eqref{eq:2} for
two rather large $p_T$ values, in analogy to Fig.~1 in
Ref.~\cite{Laenen:2000de}.  In agreement with
Fig.~\ref{fig:ptspectra_qq}, we observe a small enhancement over the
threshold-resummed result, in particular at very large $p_T$.  The
peak of the distribution is at relatively soft values of $Q_T$, a few
percent of the hard scale $m_T$.  For $p_T = 500$ GeV $Q_T$ reaches
$\bar\mu$ before the bound in Eq.~(\ref{eq:1}), so that there is no
divergence for larger $Q_T$. For $p_T=900$ GeV there is a region of $Q_T$
phase space where the bound in Eq.~(\ref{eq:1}) is saturated and the
cross section diverges.  The onset of that singularity is visible in
Fig.~\ref{fig:profile_qq}, but because of our choice $\bar\mu = 200$
GeV it does contribute a small amount to the $K$-factor in
Fig.~\ref{fig:ptspectra_qq} at the largest $p_T$ value.  We also
observe that, in case of top production, for $p_T \lesssim 863.7$ GeV, the
disk $D_2$ in eq.~(\ref{eq:3}) is fully contained in the disk $D_1$
which represents the bound in Eq.~(\ref{eq:1}). Therefore below that
value, which includes all $p_T$ values accessible experimentally, all
the enhancement of the joint resummed cross section with respect with
the threshold resummed is to be ascribed to recoil effects.
\begin{figure}[htbp]
  \centering
  \epsfig{file=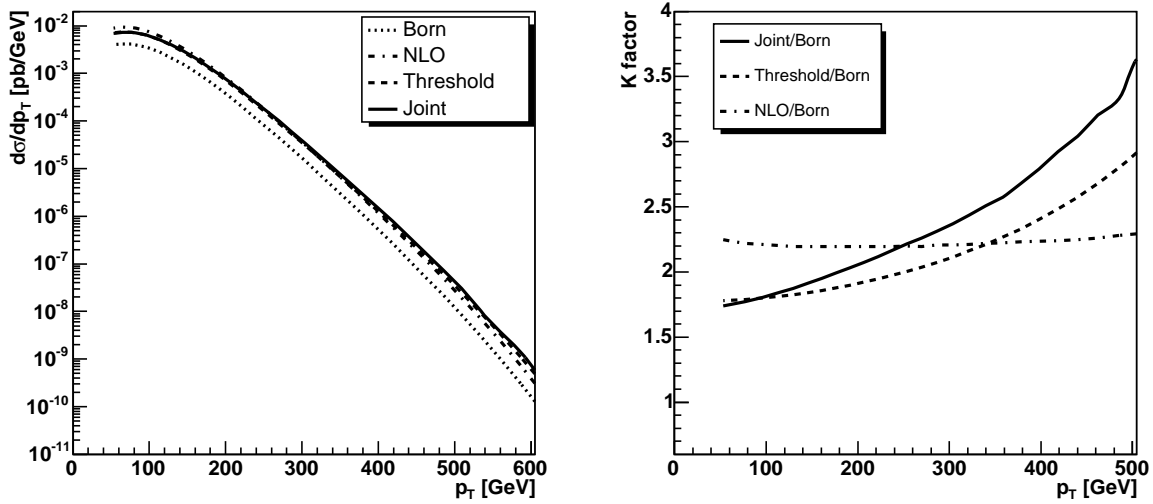,width=\textwidth}
  \caption{Top quark $p_T$ spectra and corresponding
    $K$-factors for the sum of color-singlet and color-octet $gg$ channels.
    \label{fig:ptspectra_gg}}
\end{figure}
\begin{figure}[htbp]
  \centering
  \epsfig{file=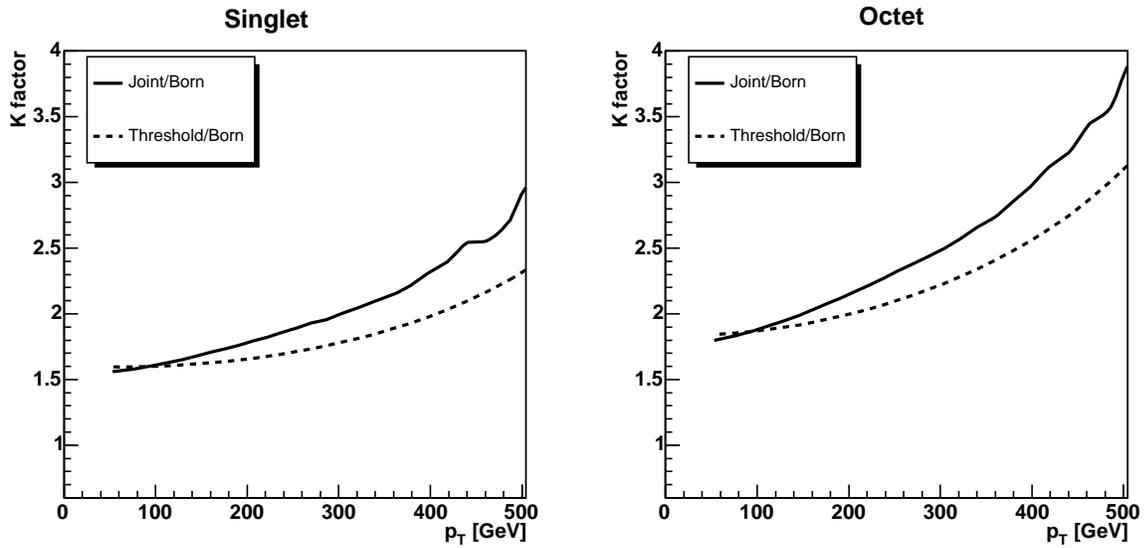,width=\textwidth}
  \caption{$K$-factors for the incoming color-singlet and color-octet 
    $gg$ channels.
    \label{fig:ptspectra_ggK}}
\end{figure}

In Figs.~\ref{fig:ptspectra_gg} we show results analogous to
Fig.~\ref{fig:ptspectra_qq}, now for the $gg$ channel summed over the
two color states, while in Fig.~\ref{fig:ptspectra_ggK} we show the
K-factors per color state.  The K-factors show a significant
enhancement for joint resummation with respect to threshold
resummation, increasing with increasing $p_T$, and strongest for octet
production. The resummed spectra are harder than the fixed order ones.
The increased enhancement of the $gg$ channel over the $q\bar q$
channel is due to the larger value for $A_g$ compared to $A_q$ in
Eq.~\eqref{eq:Eab}.

We have verified that both resummed cross sections have a reduced
scale dependence with respect to LO, as expected on general grounds
\cite{Oderda:1999im,Sterman:2000pu}.  The reduction in the resummed
cross section is somewhat less than in the exact NLO cross section,
suggesting that a matched NLL-NLO calculation would do better still in
this regard. Doing such a calculation is however beyond the scope of
our paper.

In order to assess the relevance of the last term on the right hand
side in Eq.~\eqref{eq:soft-anomalous-2}, which encodes the differences
between contributions from octet and singlet soft anomalous
dimensions, we computed the $p_T$ spectrum for the $q\bar q$ channel
with and without this term.  We found an effect ranging from 8\%
enhancement with this term included at lower $p_T$ to about 10\% 
at higher $p_T$. For the
octet $gg$ channel we found the enhancement over the singlet-only to
be 12 to 15\%.

\begin{figure}[htbp]
  \centering
  \epsfig{file=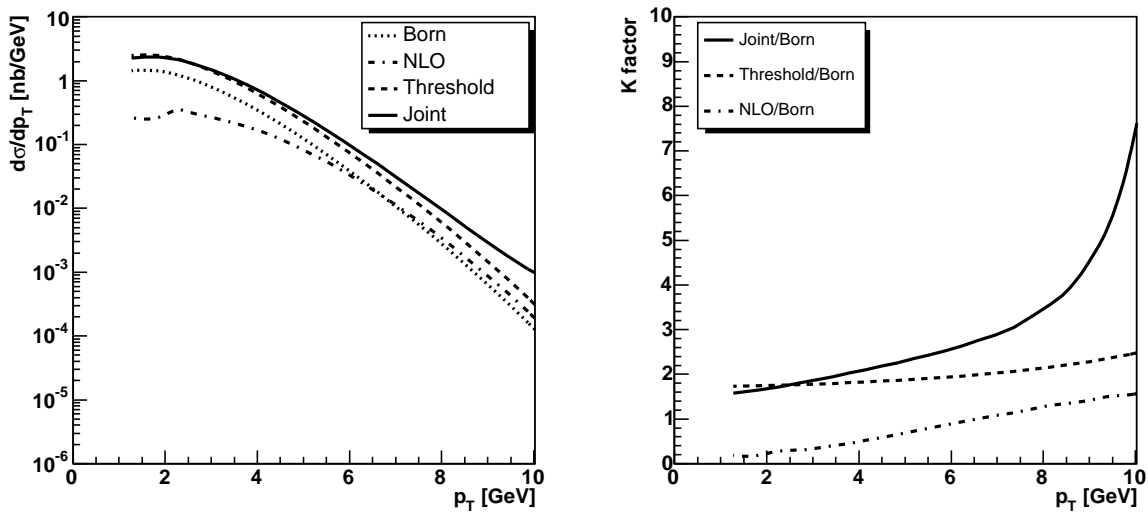,width=\textwidth}
  \caption{Bottom quark $p_T$ spectra and 
    $K$-factors for the $q\bar q$ channel.  \label{fig:ptspectra_qq_b}}
\end{figure}

\begin{figure}[htbp]
  \centering
  \epsfig{file=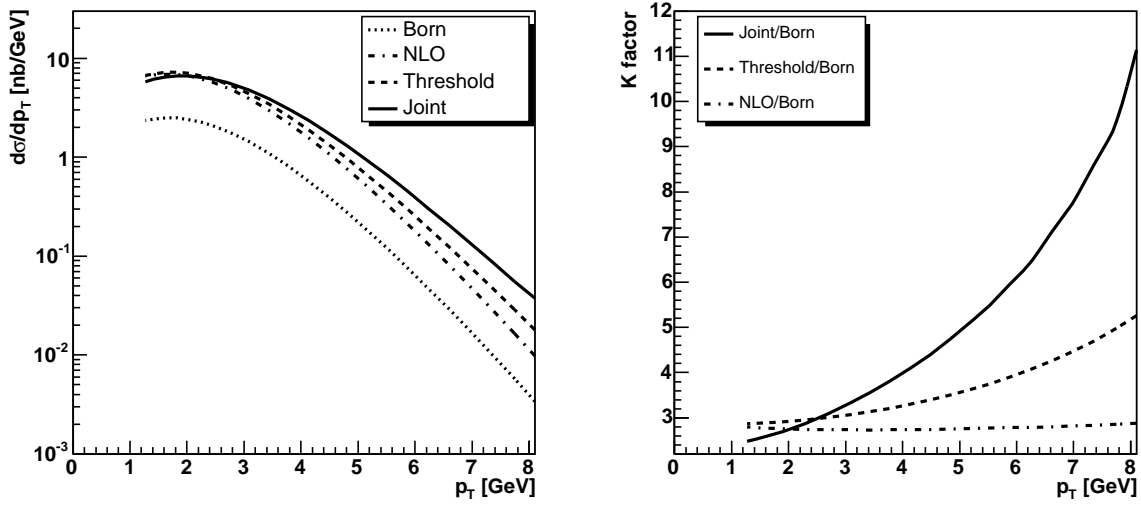,width=\textwidth}
  \caption{Bottom quark $p_T$ spectra and 
    $K$-factors for the $gg$ channel.  \label{fig:ptspectra_gg_b}}
\end{figure}

Turning now to the bottom quark $p_T$ spectrum, we took as kinematic
conditions, for illustrative purposes, $pp$ collisions at
the HERA-B cm energy ($41.6$ GeV). In these settings the
gluon channel is more dominant, and effects of joint resummation
are more noticeable than for top production in the above.
Fig.~\ref{fig:ptspectra_qq_b} is analogous to Fig.~\ref{fig:ptspectra_qq}
but the effect of joint resummation are stronger now, which holds
as well for the $gg$ channel in Fig.~\ref{fig:ptspectra_gg_b}.
\begin{figure}[htbp]
  \centering
  \epsfig{file=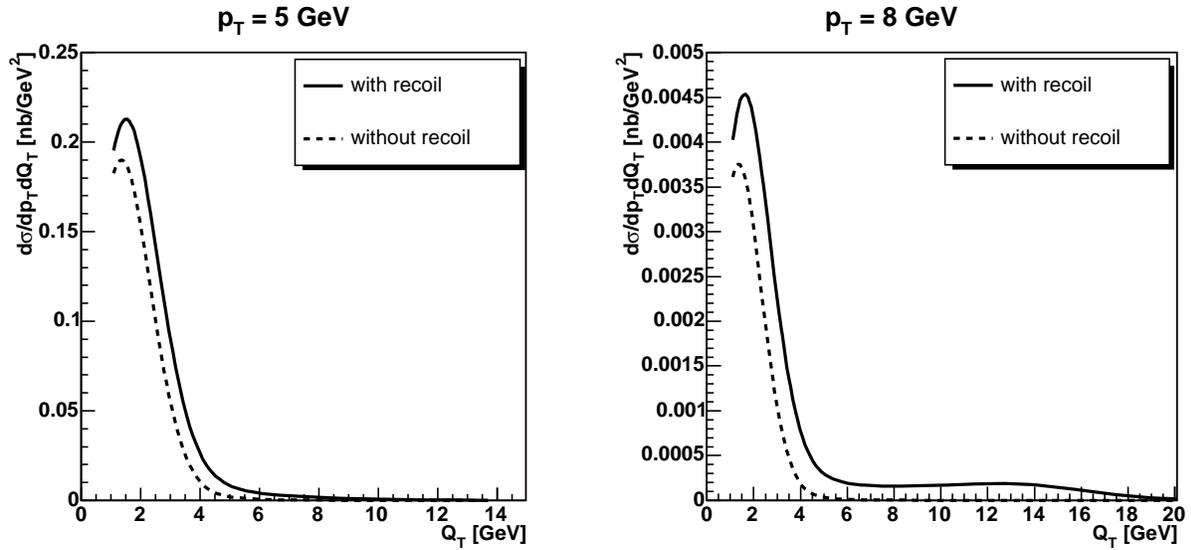,width=\textwidth}
  \caption{Bottom quark profile functions (Eq.~\eqref{eq:2})
 for the $gg$ channel, at $p_T = 5$ and  $8$ GeV. \label{fig:profile_gg_b}}
\end{figure}
In Fig.~\ref{fig:ptspectra_qq_b}, for the $q\bar{q}$ channel a significant
effect at lower $p_T$ of virtual contributions in the 
NLO cross sections, not present in the resummed versions is visible,
which is much less in the $gg$ channel. The $K$-factors are large (which
is not unusual \cite{Kidonakis:2004qe,Bonciani:1998vc,Kidonakis:2001nj} 
for this kinematical configuration), with a clear
distinction between threshold and joint resummation. The $Q_T$ profile
of the enhancement at two values of $p_T$ is shown in Fig.~\ref{fig:profile_gg_b}.
Here is there no divergence, because the bound Eq.~\eqref{eq:1} is simply not
reached.

\section{Conclusions}
\label{sec:conclusions}

In this paper we derived the joint threshold and recoil resummed heavy
quark transverse momentum distributions, to NLL accuracy. As a
corollary we obtained the pure threshold resummed result as well.
The resummed distributions contain two mutually incoherent components, 
associated with different color states (singlet and octet)
of the heavy quark pair. 
These components differ by their underlying Born process
and by anomalous dimensions governing the pattern of
wide-angle soft emission. We provided a probabilistic
interpretation of these anomalous dimensions. 

For top quark production at the Tevatron, and bottom quark production
at HERA-B, we studied the dependence of these results on the color
state, and the difference with threshold-resummation.  We found that
joint resummation gives results noticeably different from threshold
resummation, and that for a realistic range of transverse momentum the
kinematic singularity present in the formalism, which is very visible
in applications to massless (e.g. prompt photon) particle $p_T$
spectra, is well-screened by the heavy quark mass. 
Not surprisingly, resummation effects are
larger in the $gg$ channel than the $q\bar{q}$ channel, and more
noticeable for bottom-quark spectra at HERA-B than top-quark spectra
at the Tevatron. 
The effect of the pure octet part of the anomalous dimension is noticeable.
We hope that this study, taken together with
recent other ones \cite{Kulesza:2002rh, Kulesza:2003wn,Sterman:2004yk}
contributes to developing joint resummation as a viable framework
for making QCD predictions.

\subsection*{Acknowledgments}

This work was supported by the Foundation for Fundamental Research of
Matter (FOM) and the National Organization for Scientific Research
(NWO).  We are also grateful to Andreas Vogt for support in the use of the
evolution code of Ref.~\cite{Vogt:2004ns}.

\end{document}